\documentclass[aps,prl,reprint,groupedaddress,showpacs]{revtex4-1}

\usepackage{graphicx} 
\usepackage[colorlinks=true,citecolor=blue, linkcolor=blue, urlcolor=blue]{hyperref} 

\begin{document}

\title{Observing the Onset of Effective Mass}

\author{Rockson Chang}
\email[]{rchang@physics.utoronto.ca}
\author{Shreyas Potnis}
\author{Ramon Ramos}
\author{Chao Zhuang}
\author{Matin Hallaji}
\author{Alex Hayat}
\author{Federico Duque-Gomez}
\author{J. E. Sipe}
\author{Aephraim M. Steinberg}
\affiliation{Department of Physics and Institute of Optics, University of Toronto, 60 St. George Street, Toronto, ON, M5S 1A7, Canada}
\date{\today}

\begin{abstract}
The response of a particle in a periodic potential to an applied force is commonly described by an effective mass which accounts for the detailed interaction between the particle and the surrounding potential.  Using a Bose-Einstein condensate of $^{87}$Rb atoms initially in the ground band of an optical lattice, we experimentally show that the initial response of a particle to an applied force is in fact characterized by the bare mass.  Subsequently, the particle response undergoes rapid oscillations and only over timescales long compared to that of the interband dynamics is the effective mass observed to be an appropriate description.  
\end{abstract}


\maketitle


The concept of the effective mass is ubiquitous in solid state physics, allowing for a simple semiclassical treatment of the response of a particle in a solid to an external force.  The complex interaction between the particle and the surrounding potential dresses the particle with an effective mass, distinctly different from its bare mass, and allows for a description of the particles dynamics based on Newton's second law\cite{AshcroftMermin}:     
\begin{equation}
\langle a \rangle = \frac{F}{m^{\ast}_N(k)},
\label{eq:mStarThm}
\end{equation}
where $\langle a \rangle$ is the expectation value of the acceleration of the particle under an applied force $F$, and $m^{\ast}_N(k)$ is the effective mass for a particle with crystal momentum $k$ and band index $N$.  The effective mass is inversely related to the curvature of the dispersion relation, and in 1D is given by
\begin{equation}
m^{\ast}_N(k)=\hbar^2\left[ \frac{d^2}{dk^2}E_N(k)\right]^{-1},
\end{equation}
where $E_N(k)$ is the energy of the state, and $\hbar$ is Planck's constant.  The modern description of electronic conduction in solids is intimately tied to the concept of the effective mass.

\begin{figure}
\centering
\includegraphics[width=1\columnwidth]{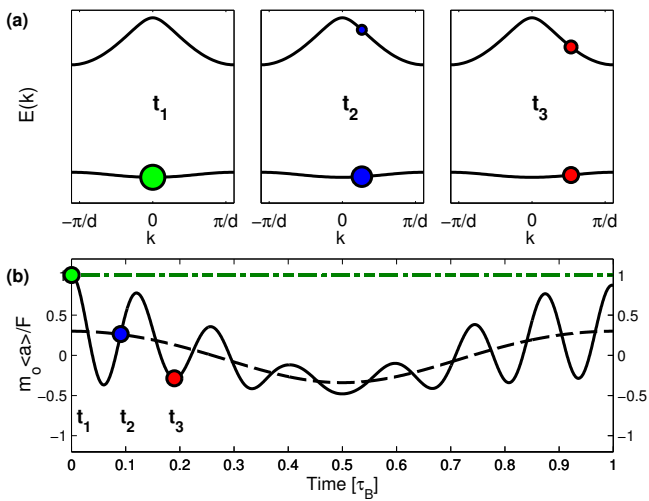}
\caption{\label{fig:illustration} Illustration of effective mass dynamics.  (a) Lattice dispersion relation showing the time evolution of an initially single-band wavepacket under an abruptly applied force.  (b) The corresponding acceleration $\langle a \rangle$, normalized by the applied force $F$.  The force results in an interaction energy with both intraband and interband contributions.  The response according to the intraband contribution alone is characterized by the lattice effective mass $m^{\ast}$ and leads to Bloch oscillations (dashed black line).  The interband contribution coherently couples the ground band to the excited bands.  This modifies the dynamics such that the total acceleration (black solid line) is initially that of a free particle with bare mass $m_0$ (bare mass response shown by green dash-dotted line), and then rapidly oscillates around the usual effective mass behaviour.}
\end{figure}

However, a direct application of Ehrenfest's theorem \cite{PS54} shows that, for a particle originally in one band, the initial acceleration due to an applied force is $F/m_0$, where $m_0$ is the bare mass, and not $F/m^{\ast}$.  This is because the external force unavoidably leads to interaction energies associated with both intraband and interband dynamics, and while the intraband portion of the interaction alone would lead to a response described by the effective mass, the additional interband contribution ensures an initial response given by the bare mass \cite{Hess88, DuqueGomez12}.  Over time, the interband coupling results in rapid oscillations in the complex amplitudes of the initial and neighbouring bands, and an acceleration which itself oscillates around $F/m^{\ast}$ (see Fig.~\ref{fig:illustration}).  In the presence of interband dephasing these oscillations die out.  The steady state however contains small contributions from neighbouring bands, as imposed by the force, such that the total acceleration tends to $F/m^{\ast}$ after the decay of the transients \cite{Adams56}.  We use the term \emph{dynamical mass} to refer to the mass associated with this transient response of the particle, and \emph{effective mass dynamics} to refer to its variation in time.  See Supplementary Information for further details on the theoretical description.

In typical solid state systems, the fast timescales of the transient oscillations and dephasing effects have thus far prohibited observation of the effective mass dynamics.  Duque-Gomez and Sipe \cite{DuqueGomez12} have recently revisited this idea specifically in the context of ultracold atoms in optical lattices, where the narrow momentum widths and inherent length and time scales involved make observation of long-range quantum coherent phenomena experimentally accessible.  In particular, the ability to switch the lattice potential faster than the characteristic response time of an atom in that lattice allows for direct observation of the momentum distribution, and has no analog in solid state experiments.  Ultracold atoms have been used extensively to simulate condensed matter phenomena such as Bloch oscillations \cite{Dahan96, OMorsch01, Wilkinson96} and quantum transport \cite{Billy08, Roati08, Brantut12, Stadler13}.  

In this letter we report on the first observation of the effective mass dynamics of an ultracold cloud of atoms in an optical lattice.  By studying the particle's response to an abruptly applied force, we show that the initial response is characterized by the bare mass and is indifferent to the presence of the lattice.  Subsequently, the particle response exhibits rapid oscillations at the bandgap frequency, and slow Bloch oscillations which are characterized by the usual effective mass.

We perform our experiment with a Bose-Einstein condensate of $^{87}$Rb prepared in a hybrid optical and magnetic trap \cite{Lin09}. This trap is formed by the overlap of a single-beam optical dipole trap with a quadruople magnietic trap.  The resulting potential has cylindrical symmetry with radial and axial harmonic trapping frequencies $f_r=$ 80 Hz and $f_z=$ 20 Hz, respectively.  In this trap we produce nearly pure condensates of about $10^5$ atoms, providing us with the long-range spatial coherence necessary for observing matter-wave phenomena.  We then ramp up a laser standing-wave pattern over 100 ms, adiabatically loading our atoms into the ground state of a 1D optical lattice with lattice constant $d=\lambda/2=$ 532 nm.  The total potential has the form
\begin{equation}
U=U_L\cos^2(k_rz)-F(t)z,
\end{equation}
where $U_L$ is the lattice depth, and $F(t)$ the applied force (initially zero).  The photon wavevector $k_r=2\pi/\lambda$ sets characteristic momentum and energy scales, $\hbar k_r$ and $E_r=\hbar^2 k_r^2/2m_0$, respectively.  We express the lattice depth in terms of the recoil energy via the dimensionless parameter $s=U_L/E_r$.  The lattice potential generated in this way is essentially defect-free, and thus avoids the complications due to scattering that arise in typical solid state systems.  The peak atomic density in the lattice is less than 3$\cdot10^{13}$ atoms/cm$^{3}$.  In this regime inter-particle interactions have been shown to be negligible up to a minor correction to the lattice depth \cite{OMorsch01}.  We estimate the lattice depth based on the momentum sideband amplitude for the $k=$ 0 Bloch state.  Comparing to a simulation of the Gross-Pitaevskii equation, we find that the dynamics occur as if the lattice had a depth $U_{eff}<U_L$.  Our results are thus compared to a single-particle analysis at lattice depth $U_{eff}$, which for our typical densities, represents a correction of less than 10\%.  For our typical lattice depths and forces we can neglect interband Landau-Zener tunnelling \cite{Bharucha97}.

To initiate the effective mass dynamics, we abruptly shift the center of the magnetic trap by up to 1 mm.  This shift is performed in 20 $\mu$s, limited by the inductance of the coils used to generate the shift, and is much faster than the typical timescale corresponding to the bandgaps in our experiment ($h/\Delta\sim$ 100 $\mu$s).  The shifted potential exerts a force $F$ on our atoms along $z$ that is essentially spatially uniform over the extent of our 20 $\mu$m sample.  After a variable evolution time in the lattice we abruptly switch off all trapping potentials, freezing out the lattice dynamics, and perform a 20 ms time-of-flight (TOF) expansion before imaging.  The effective mass dynamics result in oscillations of the average acceleration $\langle a \rangle$, and thus the average velocity $\langle v \rangle$.  The abrupt turn-off of the lattice preserves the momentum distribution, which we measure by imaging after TOF.  

\begin{figure}
\centering
\includegraphics[width=1\columnwidth]{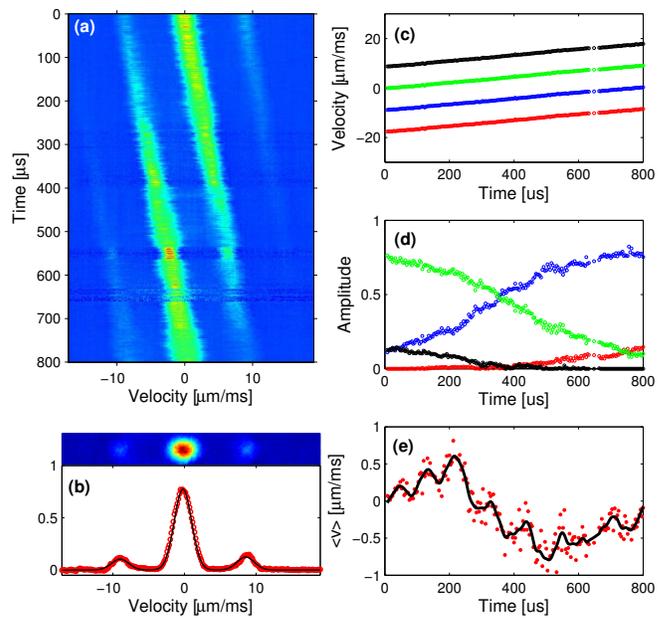}
\caption{\label{fig:data} Observation of effective mass dynamics.  $s=$ 9.4 and $F/m_0=$ 11.7 $\mu$m/ms$^2$.  (a) Composite of raw absorption images.  Each row is a slice of an absorption image taken after variable evolution time in the lattice (4 $\mu$s resolution) and 20 ms TOF, and is thus representative of the momentum distribution inside the lattice.  (b) Example absorption image (upper), and 1D profile and fit (lower).  The velocity and amplitude of each peak are extracted and plotted in (c) and (d) respectively.  (e) The average velocity is reconstructed by the sum of the velocity components weighted by their amplitudes.  The solid black curve is the result of low-pass filtering this data, and serves as a guide to the eye.}
\end{figure}

Figure~\ref{fig:data} shows the results of a typical experimental run.  The free-space momentum distribution (Fig.~\hyperref[fig:data]{\color{blue}\ref{fig:data}(a,b)}) is a diffraction pattern consisting of components separated by the recoil momentum $2\hbar k_r$ (the recoil velocity is $v_r=\hbar k_r/m_0=$ 4.3 $\mu$m/ms).  The amplitude and velocity of each peak are extracted from a fit of 4 equally-spaced, equal-width gaussians (Fig.~\hyperref[fig:data]{\color{blue}\ref{fig:data}(c,d)}).  The average velocity of the particles in the lattice as a function of time is then reconstructed from a weighted sum of these peaks (Fig.~\hyperref[fig:data]{\color{blue}\ref{fig:data}(e)}).  For each diffracted order, the momentum grows linearly in time while changing in amplitude as the particle travels across the Brillouin zone.  As one component at $-k$ rises, a component at $+k$ falls, resulting in a periodic modulation in the average velocity.    Our data show a clear oscillation in the average velocity on a millisecond timescale, consistent with Bloch oscillations \cite{Dahan96,OMorsch01,Leo92, Pertsch99}.  This phenomenon arises due to the long-range, inter-well coherence of the Bloch states and occurs at a frequency $\omega_B=Fd/\hbar$ \cite{Bloch29, Zener34}.  In addition to the Bloch oscillation we observe much faster dynamics on a 100 $\mu$s timescale, consistent with an oscillation at the bandgap frequency.  This manifests itself as a modulation of the relative amplitudes of the diffracted momentum components, as expected for an interband phenomenon.  Under the abruptly applied force, the initially single-band state will acquire amplitudes in adjacent bands over time.  The coupling to these additional bands provides contributions to the average acceleration that will oscillate at the respective band energy difference.  For narrow momentum width wavepackets and times shortly after the force is applied, the acceleration is given by \cite{Hess88, DuqueGomez12}
\begin{equation}
\langle a(t) \rangle = \frac{F}{m_0} \left[ \frac{m_0}{m^{\ast}_N}+\sum_{n\neq N}\frac{2}{m_0}\frac{p_{nN}^2}{\Delta_{nN}}\cos(\Delta_{nN}t/\hbar) \right],
\label{eq:mStarOsc}
\end{equation}
where $\Delta_{nN}$ and $p_{nN}$ are the energy gap and momentum matrix element between Bloch states in bands $n$ and $N$, respectively.  At $t=$ 0 the contributions to $\langle a \rangle$ are in phase and the initial acceleration is that of a particle with bare mass, as expected from Ehrenfest's theorem, and can be seen in Eq.~(\ref{eq:mStarOsc}) by applying the effective mass sum rule \cite{Hess88}.  For a particle initially in the ground band ($N=$ 1), the coupling is primarily to the first excited band ($n=$ 2), and the effective mass dynamics in this 2-band case are governed entirely by the bandgap $\Delta_{21}$.  As the particle traverses the Brillouin zone during a Bloch cycle, the bandgap changes, resulting in a variation in the amplitude and frequency of the effective mass oscillation.  

To quantitatively study the dynamics, we fit the the average velocity to the sum of two sinusoids, explicitly separating the Bloch oscillation from the effective mass dynamics:
\begin{equation}
v(t)=A_d\sin(\omega_dt+\phi_d)+A_B\sin(\omega_Bt+\phi_B),
\label{eq:fit}
\end{equation}
where $A$ is the amplitude, $\omega$ the frequency, and $\phi$ the phase of the oscillation.  The subscripts $d$ and $B$ indicate fitting parameters for the effective mass and Bloch oscillation, respectively.  Due to the variation of the bandgap as the particle traverses the Brillouin zone, the fitting function used is not strictly correct at times comparable to the Bloch period.  This fitting function is a compromise between capturing as many of the features of the dynamics as possible while still obtaining reliable fits.  When extracting the effective mass oscillation we fit only the first 300 $\mu$s of data (roughly 3 periods of the fast oscillation).  The choice of 300 $\mu$s reduces the fitting error due to this variation in bandgap, while providing a sufficient number of cycles to obtain an accurate estimate of the effective mass dynamics near the center of the band at $k=$ 0.

\begin{figure}
\centering
\includegraphics[width=1\columnwidth]{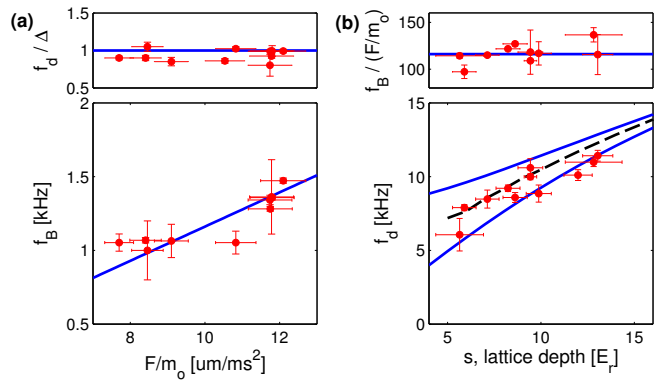}
\caption{\label{fig:timescales} (a) Variation of dynamical mass frequency (upper) and Bloch frequency (lower) with applied force.  The dynamical mass frequency, normalized by the bandgap $\Delta$, is expected to be independent of the force.  The Bloch frequency scales linearly with the applied force.  (b) Variation in Bloch frequency (upper) and dynamical mass frequency (lower) with lattice depth.  The Bloch frequency, normalized by the force divided by the bare mass $F/m_0$, is expected to be a constant $m_0d/h$ (blue solid line). The dynamical mass frequency is compared to the bandgap at $k=$ 0 and $k=k_r$ (upper and lower solid blue lines, respectively).  The dashed black curve is a comparison to simulated data fitted to Eq.~\ref{eq:fit} (see text).}
\end{figure}

Figure~\ref{fig:timescales} plots the dependence of these timescales on the applied force $F$ and lattice depth $s$.  The frequency of the slow oscillation is observed to scale linearly with the applied force (Fig.~\hyperref[fig:timescales]{\color{blue}\ref{fig:timescales}(a)}, lower),  as expected for a Bloch oscillation.  We also plot this frequency, scaled by the applied force, against lattice depth (Fig.~\hyperref[fig:timescales]{\color{blue}\ref{fig:timescales}(b)}, upper) to show that the frequency is independent of lattice depth.  The fast oscillation increases with lattice depth (Fig.~\hyperref[fig:timescales]{\color{blue}\ref{fig:timescales}(b)}, lower) and thus the bandgap, while being independent of applied force (Fig.~\hyperref[fig:timescales]{\color{blue}\ref{fig:timescales}(a)} upper).  The fitted frequency is compared to the numerically calculated bandgap at $k=$ 0 and $k=k_r$, representing the range of frequencies the particle samples as it undergoes a complete Bloch oscillation.  A more direct comparison to the data is made by fitting Eq.~(\ref{eq:fit}) to the first 300 $\mu$s of data generated from a Gross-Pitaevskii equation simulation, in the same way as we fit to the experimental data.  

\begin{figure}
\centering
\includegraphics[width=0.85\columnwidth]{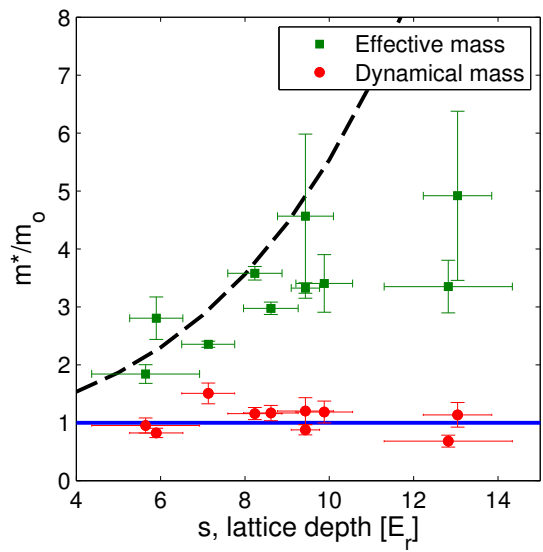}
\caption{\label{fig:mStar}  Effective mass and dynamical mass results.  The steady-state effective mass at the center of the ground band ($N=$ 1, $k=$ 0) associated with the Bloch oscillation is plotted as green squares.  As the lattice depth increases, the band flattens, and the effective mass is expected to increase (dashed black line).  The initial ($t=$ 0) dynamical mass is extracted from the full dynamics (red circles), and is expected to be consistent with the bare mass response $m^{\ast}/m_0=$ 1, independent of lattice depth (solid blue line).}
\end{figure}

From the average velocity fits we extract the initial response of the particle in the $k=$ 0 state to an applied force.  The external force is applied within 20 $\mu$s after a 20 $\mu$s delay.  This delay time is accounted for by the phase in the sinusoids of our fitting function.  This phase is negligible for the Bloch oscillation but not for the effective mass dynamics.  The initial response is evaluated at the point in time $t_0=\phi_d/\omega_d$, when the phase of the fast oscillation is zero.  The effective mass theorem Eq.~(\ref{eq:mStarThm}) describes the response of a particle to a force over timescales long compared to the interband dynamics; thus we estimate this effective mass from the Bloch oscillation alone (green squares in Fig.~\ref{fig:mStar}).  Since the band curvature decreases with increasing lattice depth, the effective mass increases.  However, the full response contains contributions from both the ``single-band" Bloch oscillation and the ``multi-band" effective mass dynamics.  The dynamical mass is estimated from the sum of these contributions, and at $t_0$ is observed to be consistent with the bare mass, independent of lattice depth (red circles in Fig.~\ref{fig:mStar}).  In deep lattices ($s>$ 10), the deduced effective mass begins to deviate from the predictions.  This is due to the growth of high-order diffraction peaks in the momentum distribution which lie beyond our imaging window.  Neglecting these peaks causes us to overestimate the amplitude of the Bloch oscillation, and thus underestimate the effective mass.  Note that this has a minimal impact on the estimate of the dynamical mass since the dominant contribution in the deep lattice regime is from the effective mass dynamics.  The error bars are given by the fit uncertainties, where the main contribution in the effective mass uncertainty comes from fitting the long timescale Bloch oscillation.  In the presence of excited band decay and interband dephasing, the effective mass dynamics are expected to reduce to the behaviour described by the usual effective mass.  In our system, this can occur due to inter-particle scattering \cite{Morandotti99, Gustavsson08}.  However, within the parameter range and timescales probed, we do not expect to see dephasing of the effective mass oscillation.

The initial response at $F/m_0$ can be understood intuitively from a microscopic perspective.  Consider, for example, a wavepacket with a narrow momentum width in the ground band of a deep lattice.  This wavepacket will consist of a superposition of many small wavepackets, each localized at a potential minimum of the lattice.  If an external force $F$ is applied, the contribution from the lattice to the total force on each localized wavepacket will initially be zero.  Thus the total force is simply the applied force, resulting in an initial acceleration $F/m_0$.  As the localized wavepackets move away from the potential minima they begin to feel a force from the lattice, leading to oscillations around the minima at the bandgap frequency.  See Supplementary Information for further details.

To our knowledge, the effective mass dynamics explored in this work have never before been observed, despite its initial prediction nearly 60 years ago.  Our work clarifies the role of the effective mass on short timescales, and is timely given the increasing interest in the effect of few-cycle laser pulses on solid targets.  This burgeoning field has shown promise for high harmonic generation \cite{Ghimire11} and fast control of electronic devices \cite{Schiffrin13,Schultze13}, where existing theoretical treatments have yet to take into account the full dynamics of the effective mass \cite{Ghimire12}.  

The authors are grateful to A. Jofre, M. Siercke and C. Ellenor for their work in developing the original Bose-Einstein condensation apparatus.  We acknowledge funding from NSERC and CIFAR.

\bibliography{mStar}

\end{document}


\title{Supplementary Information}
\date{}
\maketitle

\onehalfspacing


\section{Expectations} \label{S:Expectations}

We consider a wavepacket initially confined to a single band, with nonzero
Bloch state amplitudes in a region of the band characterized by an effective
mass $m^{\ast }$. \ If an external force $F$ is then applied for $t\geq 0$,
a na\"{\i}ve argument about the subsequent motion would be based on thinking
of the wavepacket as evolving according to an effective Hamiltonian 
\begin{equation}
\hat{H}_{eff}=\frac{\hat{p}^{2}}{2m^{\ast }}-\hat{x}F. \nonumber
\end{equation}
Were this valid the response would be simple, and would follow from a simple
calculation. \ Identifying the velocity operator $\hat{v},$%
\begin{equation}
\hat{v}\equiv \frac{1}{i\hbar }\left[ \hat{x},\hat{H}_{eff}\right] =\frac{%
\hat{p}}{m^{\ast }},\nonumber
\end{equation}
from Ehrenfest's theorem we would have 
\begin{equation}
\left<a(t) \right> =\frac{1}{%
i\hbar }\left\langle \psi (t) \middle| \left[ \hat{v},\hat{H}_{eff}\right] \middle| \psi
(t)\right\rangle =\frac{F}{m^{\ast }},  \label{guesswrong}
\end{equation}%
where the acceleration
\begin{equation}
	\left<a(t) \right> \equiv \frac{d}{dt}\left\langle \psi (t)|\hat{v}|\psi (t)\right\rangle \nonumber
\end{equation}
would be characterized by the effective mass. \ Here
the presence of the underlying periodic potential, which leads to an
effective mass $m^{\ast }$ differing from the bare mass $m_o$, obviously plays
a crucial but implicit role.

Yet a different simple argument conflicts with this. \ We imagine that
before an external force $F$ is applied we have a wavepacket with nonzero
Bloch state amplitudes only in the lowest band and centered about $k=0$. \
At least if the minima of the periodic potential ($V(x)=V(x+d)$, where $d$
is the lattice constant) are very deep, then this wavepacket will
essentially be the sum of a number of small, localized wavepackets, each
centered at a local potential minimum. The expectation value of the force
from the lattice on each such small, well-centered wavepacket vanishes.
Consequently when the external force $F$ is applied one could argue that any
dynamical effect of the lattice could (at least initially) be neglected; the
picture would be of each well-centered wavepacket subject to the external
force as it would be were no lattice present, and thus one would expect the
overall initial acceleration to be governed by the bare mass $m_o$,%
\begin{equation}
\left<a(t) \right>^{o}=\frac{F}{m_o},  \label{guess}
\end{equation}%
where the superscript $^{o}$ indicates the initial time when the force is
applied.

This argument could certainly be criticized as overly schematic, but its main fault is its implication that the conclusion (\ref%
{guess}) might be limited to deep lattice potentials and particular
wavepackets; in fact, the result is more general than it might seem. The expectation value of the force due to the
lattice on \textit{any }wavepacket consisting of Bloch state components from
only a single band vanishes, \textit{regardless of whether the lattice
potential is deep or not.} \ To see this, consider first a general state $
\left\vert \psi \right\rangle $ with coordinate representation $
\psi(x)=\left\langle x| \psi \right\rangle $. \ The expectation value
of the force due to the lattice is then given by 
\begin{eqnarray*}
-\left\langle \psi \middle| \frac{\partial V(\hat{x})}{\partial \hat{x}} \middle|
\psi \right\rangle  &=&-\int dx\, \psi^{\ast }(x)\left( \frac{\partial
V(x)}{\partial x}\right) \psi(x) \\
&=&\frac{1}{i\hbar }\int \left[ \left\langle \psi \middle| \hat{p} \middle| x\right\rangle \left\langle x|V(\hat{x})| \psi \right\rangle
-\left\langle \psi \middle| V(\hat{x}) \middle| x \right\rangle \left\langle x \middle| \hat{p} \middle|
\psi \right\rangle \right] dx,
\end{eqnarray*}
where of course 
\begin{equation}
\left\langle \psi \middle| \hat{p} \middle| x\right\rangle ^{\ast }=\left\langle x \middle| \hat{p} \middle| \psi \right\rangle = \frac{\hbar }{i}\frac{\partial \psi(x)}{\partial x} \nonumber
\end{equation}
and 
\begin{equation}
\left\langle x \middle| V(\hat{x}) \middle| \psi \right\rangle =V(x) \psi(x). \nonumber
\end{equation}
So 
\begin{eqnarray}
-\left\langle \psi \middle| \frac{\partial V(\hat{x})}{\partial \hat{x}} \middle| \psi \right\rangle  &=& \frac{1}{i\hbar} \left\langle \psi \middle| \left[ \hat{p},V(\hat{x})\right] \middle| \psi \right\rangle = \frac{1}{i\hbar} \left\langle \psi \middle| \left[ \hat{p},\hat{H}_{o}\right] \middle| \psi \right\rangle , \label{forceresult}
\end{eqnarray}
where 
\begin{equation}
\hat{H}_{o}=\frac{\hat{p}^{2}}{2m_o}+V(\hat{x})  \label{Hnought}
\end{equation}%
is the full lattice Hamiltonian. \ Restricting ourselves now to a wavepacket 
$\bar{\psi}(x)$ consisting solely of Bloch states within a given band $n=N$, 
\begin{equation}
\bar{\psi}(x)=\int dk\;c_{N}(k)\psi _{Nk}(x),  \label{psibar}
\end{equation}%
where $\psi _{nk}(x)=\left\langle x|\psi _{nk}\right\rangle $ are the Bloch
functions, a simple calculation gives that the expectation value of the
force due to the lattice on the wavepacket indeed vanishes, 
\begin{equation}
-\left\langle \bar{\psi} \middle| \frac{\partial V(\hat{x})}{\partial \hat{x}} \middle| \bar{%
\psi}\right\rangle =\frac{1}{i\hbar }\left\langle \bar{\psi} \middle| \left[ \hat{p},%
\hat{H}_{o}\right] \middle| \bar{\psi}\right\rangle =0  \label{noforce}
\end{equation}%
from (\ref{forceresult}), since $\hat{H}_{o}\left\vert \psi
_{nk}\right\rangle =E_{n}(k)\left\vert \psi _{nk}\right\rangle $, where $%
E_{n}(k)$ is the energy of the Bloch state of band $n$ at $k$, and 
\begin{equation}
\left\langle \psi _{nk}|\hat{p}|\psi _{mk^{\prime }}\right\rangle
=p_{nm}(k)\delta (k-k^{\prime }).  \label{pmatrix}
\end{equation}

From the perspective of condensed matter physics this is no surprise. A
standard example in that subject is a wavepacket evolving according to (\ref{Hnought}%
), where an initial wavepacket (\ref{psibar}) involving a single band would
evolve as $\bar{\psi}(x)\rightarrow \bar{\psi}(x,t)=\left\langle x|\bar{\psi}%
(t)\right\rangle $, 
\begin{equation}
\bar{\psi}(x,t)=\int dk\;\bar{c}_{N}(k,t)\psi _{Nk}(x),  \label{psibart}
\end{equation}%
with simply $\bar{c}_{N}(k,t)=c_{N}(k)\exp (-itE_{N}(k)/\hbar )$. \ The
velocity operator is given by 
\begin{equation}
\hat{v}\equiv \frac{1}{i\hbar }\left[ \hat{x},\hat{H}_{o}\right] =\frac{\hat{%
p}}{m_o}, \nonumber
\end{equation}
but because the momentum operator is diagonal in $k$ (\ref{pmatrix}), in the
expectation value $\left\langle \bar{\psi}(t)|\hat{v}|\bar{\psi}%
(t)\right\rangle $ the phase factors $\exp (-itE_{N}(k)/\hbar )$ are
irrelevant, and we have the well-known result that $\left\langle \bar{\psi}%
(t)|\hat{v}|\bar{\psi}(t)\right\rangle $ is independent of time. \ From the
perspective of Ehrenfest's theorem,%
\begin{equation}
\frac{d}{dt}\left\langle \bar{\psi}(t)|\hat{v}|\bar{\psi}(t)\right\rangle =%
\frac{1}{i\hbar m_o}\left\langle \bar{\psi}(t)|\left[ \hat{p},H_{o}\right] |%
\bar{\psi}(t)\right\rangle =0 \nonumber
\end{equation}
where the commutator term vanishes because of (\ref{noforce})\ and the fact
that $\bar{\psi}(x,t)$ (\ref{psibart}) is of the same form as $\bar{\psi}(x)$
(\ref{psibar}), $\left\langle \bar{\psi}(t)|\hat{v}|\bar{\psi}%
(t)\right\rangle $ is constant precisely because the expectation value of
the force due to the lattice on such a wavepacket vanishes, even as it
propagates and disperses.

Turning now to the imposition of an external force, where $\hat{H}_{o}\rightarrow \hat{H}$ with 
\begin{equation}
\hat{H}=\hat{H}_{o}-\hat{x}F  \label{Hfull}
\end{equation}
for $t\ge0$, we still have 
\begin{equation}
\hat{v}=\frac{\hat{p}}{m_o}  \label{vuse}
\end{equation}
and for an initial wavepacket consisting of Bloch states from a
single band $N$ (\ref{psibar}) the initial state $\left\vert \bar{\psi}%
\right\rangle $ evolves to a state $\left\vert \psi (t)\right\rangle $ more
complicated than $\left\vert \bar{\psi}(t)\right\rangle $; still, from
Ehrenfest's theorem at $t=0$ we have 
\begin{eqnarray}
\left< a(t) \right>^o &=& \frac{1}{i\hbar m_o}\left\langle \bar{\psi} \middle| \left[ \hat{p},H%
\right] \middle| \bar{\psi}\right\rangle =\frac{1}{i\hbar m_o}\left\langle \bar{\psi} \middle| \left[ \hat{p},H_{o}\right] \middle|%
\bar{\psi}\right\rangle -\frac{F}{i\hbar m_o}\left\langle \bar{\psi} \middle| \left[ 
\hat{p},\hat{x}\right] \middle| \bar{\psi}\right\rangle  \label{rightanswer} \\
&=&-\frac{F}{i\hbar m_o}\left\langle \bar{\psi} \middle| \left[ \hat{p},\hat{x}\right] \middle|%
\bar{\psi}\right\rangle =\frac{F}{m_o}  \nonumber
\end{eqnarray}%
in agreement with (\ref{guess}) because as the external force $F$ is applied
the expectation value of the force of the lattice on the wavepacket indeed
vanishes. \ This is the argument of Pfirsch and Spenke\cite{Pfirsch54}.  So the
intuition leading to (\ref{guess}) is correct, and that to (\ref{guesswrong}%
) is incorrect.


\section{Position operators} \label{S:PositionOperators}

Nonetheless, it would be harsh to completely dismiss the argument leading to
(\ref{guesswrong}), for it does capture \textit{some }of the physics, and in
a way that in fact can be made precise. \ To show this we allow for a more
general wavepacket than considered in \eqref{psibar}, and here write

\begin{equation}
\left\vert \psi \right\rangle =\sum_{n}\int dk\;c_{n}(k)\left\vert \psi
_{nk}\right\rangle ,  \label{generalket}
\end{equation}%
or 
\begin{equation}
\left\langle x|\psi \right\rangle =\psi (x)=\sum_{n}\int dk\;c_{n}(k)\psi
_{nk}(x),  \label{generalwavepacket}
\end{equation}%
and we let the sum over $n$ range over many bands. \ The integral over $k$
is taken over one Brillouin zone; we choose the $c_{n}(k)$ to be nonzero
only in this zone and away from its edges. \ Then using (\ref%
{generalwavepacket}) the expectation value of the position operator is
easily found to be 
\begin{eqnarray}
\left\langle \hat{x}\right\rangle  &=&\left\langle \psi |\hat{x}|\psi
\right\rangle   \label{xdef} = \int \psi ^{\ast }(x)x\psi (x)dx \\
&=&\frac{i}{2}\sum_{n}\int \left[ c_{n}^{\ast }(k)\frac{\partial c_{n}(k)}{%
\partial k}-\frac{\partial c_{n}^{\ast }(k)}{\partial k}c_{n}(k)\right] dk +\sum_{n,m}\int dk\;\xi _{nm}(k)c_{n}^{\ast }(k)c_{m}(k).  \nonumber
\end{eqnarray}%
The $\xi _{nm}(k)$ are the Lax connections\cite{Lax74} relating the periodic parts
of the Bloch functions,%
\begin{equation}
i\frac{\partial u_{nk}(x)}{\partial k}=\sum_{m}u_{mk}(x)\xi _{mn}(k),
\label{Lax}
\end{equation}%
where we write the Bloch functions as 
\begin{equation}
\psi _{nk}(x)=\frac{u_{nk}(x)}{\sqrt{2\pi }}e^{ikx}, \nonumber
\end{equation}
with $u_{nk}(x+d)=u_{nk}(x)$. \ It is natural to write (\ref{xdef}) as 
\begin{equation}
\left\langle \hat{x}\right\rangle =\left\langle \hat{x}\right\rangle
_{intra}+\left\langle \hat{x}\right\rangle _{inter}, \nonumber
\end{equation}
where 
\begin{eqnarray*}
\left\langle \hat{x}\right\rangle _{intra} &=&\frac{i}{2}\sum_{n}\int \left[
c_{n}^{\ast }(k)\frac{\partial c_{n}(k)}{\partial k}-\frac{\partial
c_{n}^{\ast }(k)}{\partial k}c_{n}(k)\right] dk \\
&&+\sum_{n}\int dk\;\xi _{nn}(k)c_{n}^{\ast }(k)c_{n}(k)
\end{eqnarray*}%
can be identified as an ``intraband" contribution, and 
\begin{equation}
\left\langle \hat{x}\right\rangle _{inter}=\sum_{n,m}^{\prime }\int dk\;\xi
_{nm}(k)c_{n}^{\ast }(k)c_{m}(k) \nonumber
\end{equation}
is an ``interband" contribution; the prime indicates that the terms
with $n=m$ are to be excluded. We can formally write the two components as 
\begin{equation}
\left\langle \hat{x}\right\rangle _{intra} = \sum_{n,m}\int dkdk^{\prime }c_{n}^{\ast }(k)\left\langle \psi _{nk}|\hat{%
x}_{i}|\psi _{mk^{\prime }}\right\rangle c_{m}(k^{\prime }) \nonumber
\end{equation}%
and 
\begin{equation}
\left\langle \hat{x}\right\rangle _{inter} = \sum_{n,m}\int dkdk^{\prime }c_{n}^{\ast }(k)\left\langle \psi _{nk}|\hat{%
x}_{e}|\psi _{mk^{\prime }}\right\rangle c_{m}(k^{\prime }), \nonumber
\end{equation}%
where 
\begin{eqnarray}
\left\langle \psi _{nk}|\hat{x}_{i}|\psi _{mk^{\prime }}\right\rangle 
&=&\delta _{nm}\left[ \delta (k-k^{\prime })\xi _{nn}(k)+i\frac{\partial }{%
\partial k}\delta (k-k^{\prime })\right] ,  \label{xixe} \\
\left\langle \psi _{nk}|\hat{x}_{e}|\psi _{mk^{\prime }}\right\rangle 
&=&\left( 1-\delta _{nm}\right) \delta (k-k^{\prime })\xi _{nm}(k) 
\nonumber
\end{eqnarray}%
are taken as the matrix elements that define an \emph{intraband position operator} $\hat{x}_{i}$ and an \emph{interband position operator} $\hat{x}_{e}$. \
The diagonal terms $\xi _{nn}(k)$ must appear in (\ref{Lax}) if the $%
u_{nk}(x)$ are to be well-defined over the Brillouin zone and its extension;
their appearance in (\ref{xixe}) will guarantee that terms of physical
significance are independent with respect to global (i.e., position
independent) but $k$-dependent phase factors that may be chosen to multiply
the $\psi _{nk}(x)$ [\ref{Lax74r}]. In higher dimensions the $\xi _{nn}(k)$ become vector
quantities and their curl is the Berry curvature, leading to anomalous
velocity effects \cite{Blount62}, for example. \ In 1D they are more benign. This all
harks back to the very early work by Adams and Blount \cite{Blount62}; for a more recent
presentation in this spirit and references to earlier work see Aversa and Sipe \cite{Aversa95}. \ It is
useful to define $x_{nm}(k)\equiv (1-\delta _{nm})\xi _{nm}(k)$, so that $%
x_{nm}(k)=0$ if $n=m$; the commutator $\left[ H_{o},X\right] =\hbar P/im_o$
then leads to 
\begin{equation}
x_{nm}(k)=\frac{\hbar p_{nm}(k)}{im_oE_{nm}(k)},  \label{xmatrix}
\end{equation}%
for $m\neq n$, where $E_{nm}(k)\equiv E_{n}(k)-E_{m}(k)$.

Although $\hat{x}_{e}$ is associated with interband transitions and $%
\hat{x}_{i}$ with intraband transitions (see (\ref{xixe})), and $\hat{x}=%
\hat{x}_{e}+\hat{x}_{i}$, the operators $\hat{x}_{e}$ and $\hat{x}_{i}$ do not commute \cite{Aversa95} and cannot be thought of as kinematically independent; this in itself
points out that in general a simple separation of motion into ``intraband"
and ``interband" components is not possible.

Nonetheless, if we start with a wavepacket (\ref{psibar}) initially confined
to a single band and then subject to an external force, with the dynamics
following from the Hamiltonian (\ref{Hfull}), we can consider separately the
initial effects of the ``intraband component" of the interaction, $-\hat{x}%
_{i}F$, and the ``interband component", $-\hat{x}_{e}F$, 
\begin{equation}
\left\langle a(t) \right\rangle^{o}=\left\langle a(t) \right\rangle^{o}_{\text{intra}}+\left\langle a(t) \right\rangle^{o}_{\text{inter}}.  \label{accsum}
\end{equation}%
The initial acceleration due to the intraband component of the interaction
is 
\begin{equation}
\left\langle a(t) \right\rangle^{o}_{\text{intra}} =\frac{F}{im_o\hbar }\int dkdk^{\prime }\;c_{N}^{\ast }(k)c_{N}(k^{\prime
})\left\langle \psi _{Nk}|[\hat{x}_{i},\hat{p}]|\psi _{Nk^{\prime
}}\right\rangle,   \label{accintra}
\end{equation}%
where we have used (\ref{vuse}). Analogously, for the interband component we have
\begin{equation}
\left\langle a(t) \right\rangle^{o}_{\text{inter}} = \frac{F}{im_o\hbar }\int dkdk^{\prime }\;c_{N}^{\ast }(k)c_{N}(k^{\prime
})\left\langle \psi _{Nk}|[\hat{x}_{e},\hat{p}]|\psi _{Nk^{\prime
}}\right\rangle.  \label{accinter}
\end{equation}%

The matrix elements of the commutators $[\hat{x}_{i},\hat{p}]$ and $[\hat{x}_{e},\hat{p}]$ between two Bloch states are found to be \cite{Aversa95}
\begin{equation}
	\left< \psi_{Nk} \middle| \left[ \hat{x}_{i} , \hat{p} \right] \middle| \psi_{Nk'} \right> =  \frac{i \hbar m_o}{m^{\ast}_{N}(k)} \delta(k-k'), \label{E:IntraBloch}
\end{equation}
where $m^{\ast}_{N}(k)$ is the usual effective mass for band $N$, and
\begin{equation}
	\left< \psi_{Nk} \middle| \left[ \hat{x}_{e} , \hat{p} \right] \middle| \psi_{Nk'} \right> =  i \hbar \left( 1 -  \frac{m_o}{m^{\ast}_N(k)} \right) \delta(k-k'),\label{E:InterBloch}
\end{equation}
which follows immediately from \eqref{E:IntraBloch} and the fact that $[\hat{x},\hat{p}]=i\hbar$, with $\hat{x} = \hat{x}_i + \hat{x}_e $. Using  \eqref{E:IntraBloch} in \eqref{accintra}, we find 
\begin{equation}
	\left\langle a(t) \right\rangle^{o}_{\text{intra}} = \left( \int dk\;\frac{1}{m_{N}^{\ast }(k)}\left\vert c_{N}(k)\right\vert^{2}\right) F.
\end{equation}
That is, if the wavepacket responded to \textit{only} the intraband part of
the interaction, it would respond with the effective mass (appropriately
averaged over the wavepacket); this makes precise the intuition that led to (%
\ref{guesswrong}).

For the full initial acceleration, however, we must also consider the
interband part of the interaction. Combining the two contributions in \eqref{accsum}, we have 
\begin{equation}
	\left< a(t) \right>^o = \left( \int dk \frac{1}{m_o} |c_N(k)|^2 \right) F = \frac{F}{m_o}.
\end{equation} 
We have recovered (\ref{rightanswer}), and shown that while the intraband component
of the interaction would alone lead to a response characterized by the
effective mass, the result that the wavepacket response is characterized by
the bare mass can be understood as arising from the combined effects of the
intraband and interband components of the interaction.


\section{Modified Bloch states}

So the motion of a wavepacket subject to an external force involves in
general a complicated interplay of interband and intraband motion. \ How
then to push beyond the initial value of acceleration identified above, and
identify at least approximately the full $\left\vert \psi (t)\right\rangle $
equal to (\ref{psibar}) at $t=0$ and evolving henceforth according to (\ref%
{Hfull})?

Two of us earlier\cite{Duque12} addressed this problem employing the \textit{modified
Bloch states }introduced years ago by Adams and Wannier for situations where Landau-Zener tunneling can be neglected\cite{Adams56, Wannier60}. These states, which
we denote by $\left\vert \phi _{nk}\right\rangle $, are labeled by a band
index $n$ and crystal momentum $k$ as are the Bloch states, but they are not
eigenstates of the $H_{o}$; rather, they are built in such a way that a wavepacket initially of the form
\begin{equation}
	\left| \bar{\phi} \right>  = \int dk \, \bar{b}_N(k) \left \vert \phi_{N k} \right> \nonumber
\end{equation} 
evolves in time, according to the full Hamiltonian \eqref{Hfull}, as 
\begin{equation}
	\left| \bar{\phi}(t) \right>  = \int dk \, \bar{b}_N(k,t) \left \vert \phi_{N k} \right>, \nonumber
\end{equation} 
so that for $t>0$ the wavepacket is still formed by modified Bloch states with the same initial band index $N$. In other words, the wavepacket remains in the new ``modified band" in the presence of an external force. 
The amplitude $\bar{b}_N(k,t)$ is
\begin{equation}
	\bar{b}_N(k,t) \approx \bar{b}_N(\kappa)e^{-i \gamma_N(\kappa,t)} \nonumber
\end{equation}
to first order in the force, where we introduced a wavevector moving through the Brillouin zone, $\kappa \equiv k-Ft/\hbar$, and a phase associated with the energy of band $N$ renormalized by the diagonal part of the Lax connection, 
\begin{equation}
	\gamma_N(\kappa,t) \equiv \frac{1}{\hbar} \int_{0}^{t} (E_N(\kappa+Ft'/\hbar)-F\xi_{NN}(\kappa+Ft'/\hbar)) dt'. \nonumber
\end{equation}

The modified Bloch states can be constructed from the usual Bloch states by a unitary transformation\cite{Wannier60}. To first order in the force we have
\begin{equation}
	\left\vert \phi _{nk}\right\rangle \approx \left\vert \psi _{nk}\right\rangle +\sum_{n'} \left\vert \psi _{n'k}\right\rangle \frac{F x_{n'n}(k)}{E_{n'n}(k)}, \nonumber
\end{equation}
which clearly shows that the wavepacket $\left| \bar{\phi}(t) \right>$ is formed by a superposition of Bloch states where the largest contribution comes from the original band $N$ (associated with $\left| \psi_{N k} \right>$). The amplitudes in the other bands are essential to keep the wavepacket in the \emph{modified} band $N$ (associated with $\left| \phi_{N k} \right>$) for $t > 0$. Furthermore, a wavepacket of this form responds at all times with the effective mass $m_N^{\ast}(k)$,
\begin{equation}
	\frac{d}{dt} \left< \bar{\phi}(t) \middle| \hat{v} \middle| \bar{\phi}(t) \right> = \left( \int dk \, \frac{1}{m_N^{\ast}(k)} |\bar{b}_N(k,t)|^2 \right) F, \nonumber
\end{equation}
and performs Bloch oscillations as it moves through the Brillouin zone\cite{Adams56}. In this sense the Bloch oscillations can be regarded as a \emph{single band} effect with respect to the \emph{modified} bands associated with the modified Bloch states. 

Clearly, the wavepacket $\left| \bar{\phi}(t) \right>$ does not consist of states from only one of the original bands, and hence, were the force turned on at $t=0$, its response at that time would not be characterized by the bare mass as shown before for a wavepacket of the form \eqref{psibar} at $t=0$. How can we then describe the dynamics of a wavepacket initially consisting of Bloch states from only one of the original bands? Since the dynamics are easily described using the modified Bloch states, we construct our wavepacket $\left| \psi(t) \right>$ in terms of them so that $\psi(x, t=0)$ equals the $\bar{\psi}(x)$ of \eqref{psibar}. This requires the inclusion of modified Bloch states with $n \ne N$, 
\begin{equation}
\left| \psi(t) \right>  =\sum_n \int dk \, b_n(k,t) \left \vert \phi_{n k} \right>, \label{E:WapacketMod}
\end{equation}
with the $b_n(k,t=0)$ chosen to satisfy the initial condition. The amplitudes in this case are given by
\begin{equation}
	b_n(k,t) \approx \begin{cases} c_N(\kappa) e^{-i \gamma_{n}(\kappa, t)} \text{, if $n=N$} \\ -c_N(\kappa) \left( \frac{F x_{nN}(k)}{E_{nN}(k)}\right)e^{-i \gamma_{n}(\kappa, t)} \text{, if $n \ne N$} \end{cases}, \label{E:ModAmpl}
\end{equation}
to first order in the force\cite{Duque12}. Since this wavepacket is initially in the original Bloch band $N$, it will behave with the bare mass at $t=0$; at later times, however, it will not simply respond with the usual effective mass because of the presence of additional amplitudes $b_n(k,t)$ for $n \ne N$. The expectation value of the acceleration will oscillate around the effective mass behaviour. Details have been presented earlier\cite{Duque12}.


\section{Dynamics of the effective mass}

The deviations from the usual effective mass behaviour of the wavepacket \eqref{E:WapacketMod} for $t>0$ can be explained in terms of the \emph{intraband} and \emph{interband contributions} of the interaction with the force, as done in section \ref{S:PositionOperators} for the initial response. To illustrate this point we move back to the representation in terms of the usual Bloch states. From the amplitudes \eqref{E:ModAmpl} we conclude\cite{Duque12} that in this basis $\left| \psi(t) \right>$ can be written as \eqref{generalket} with $c_n(k)$ replaced by
\begin{equation}
	c_{n}(k,t) = \begin{cases} c_N(\kappa) e^{-i \gamma_{n}(\kappa, t)} \text{, if $n=N$}  \\ c_N(\kappa) \left(\frac{F x_{nN}(k)}{E_{nN}(k)} e^{-i \gamma_{N}(\kappa, t)} - \frac{F x_{nN}(\kappa)}{E_{nN}(\kappa)} e^{-i \gamma_{n}(\kappa, t)} \right) \text{, if $n \ne N$} \end{cases}. \label{E:BlochAmpl}
\end{equation}

For times $t>0$ we need to include the contributions of both commutators $\left[ \hat{p}, \hat{H}_o \right]$ and $\left[ \hat{p}, \hat{x} \right]$ when applying Ehrenfest's theorem to calculate the expectation value of the acceleration (see expression \eqref{rightanswer}). Accordingly, we write
\begin{equation}
	\left<a (t) \right> = \sum_{n,m} \iint  \frac{dk dk'}{i \hbar m_o} c_n^{\ast}(k,t) c_m(k',t) \left( \left< \psi_{nk} \middle| \left[ \hat{p}, \hat{H}_o \right] \middle| \psi_{mk'} \right> + F \left< \psi_{nk} \middle| \left[ \hat{x} , \hat{p} \right] \middle| \psi_{mk'} \right> \right). \label{E:Ehrenfest}
\end{equation}

As discussed in section \ref{S:Expectations}, the first commutator in this expression is nonzero only for $n \ne m$,
\begin{equation}
	 \left< \psi_{nk} \middle| \left[ \hat{p} , \hat{H}_o \right] \middle| \psi_{mk'} \right> =E_{nm}(k) p_{nm}(k) \delta(k-k'). \nonumber
\end{equation}
The second commutator in \eqref{E:Ehrenfest} is also proportional to $\delta(k-k')$ and its diagonal terms (with respect to the band indices) are related to the effective mass, as shown in \eqref{E:IntraBloch} and \eqref{E:InterBloch}. The commutation relation $[\hat{x},\hat{p}]=i\hbar$ makes the off-diagonal components vanish,
\begin{equation}
	\left< \psi_{n k} \middle| \left[ \hat{x}, \hat{p}\right] \middle| \psi_{m k'} \right> = 0 \text{, for $n \ne m$}. \nonumber
\end{equation}

Using these results for the commutators in \eqref{E:Ehrenfest} and keeping terms to first order in the force, we conclude that the expectation value of the acceleration has two contributions
\begin{equation}
\left<a (t) \right> = \left<a (t) \right>_{N} + \left<a (t) \right>_{\text{coh}}. \label{E:AccelNandCoh}
\end{equation}
The first one, given by
\begin{equation}
	\left<a (t) \right>_{N} = \left<a (t) \right>_{N, \text{ intra}}  + \left<a (t) \right>_{N, \text{ inter}} , \label{E:AccelN}
\end{equation}
is associated with the population of band $N$ through an intraband term,
\begin{equation}
	\left<a (t) \right>_{N, \text{ intra}} = \left( \int dk\, \frac{1}{m_{N}^{\ast }(k)}\left\vert c_{N}(\kappa)\right\vert^{2}\right) F, \label{E:AccIntraAll}
\end{equation}
and an interband term,
\begin{equation}
	\left<a (t) \right>_{N, \text{ inter}} = \left( \int dk \left(\frac{1}{m_o} - \frac{1}{m_{N}^{\ast }(k)} \right) \left\vert c_{N}(\kappa)\right\vert^{2}\right) F. \label{E:AccInterAll}
\end{equation}
Recall that the population of band $N$ is simply $|c_N(k,t)|^2 = |c_N(\kappa)|^2$ according to \eqref{E:BlochAmpl}; thus, the two contributions in \eqref{E:AccIntraAll} and \eqref{E:AccInterAll} match $\left< a(t) \right>_{\text{intra}}^{o}$ and $\left< a (t) \right>_{\text{inter}}^{o}$ at $t=0$. 

Only at this initial time  the acceleration is given exclusively by $\left<a (t) \right>_{N}$. At later times we need to consider the second term in \eqref{E:AccelNandCoh}, which results from the coherence between band $N$ and the other bands. It can be written as
 \begin{equation}
 	\left<a (t) \right>_{\text{coh}} = \sum_{n \ne N} \int \frac{dk}{i \hbar m_o} \, c_N^{\ast}(k,t) c_n(k,t) E_{Nn}(k) p_{Nn}(k) + \text{C.C.}, \label{E:AccelCoh}
 \end{equation}  
where \emph{C.C.} denotes the complex conjugate of the first term. If we ignored $\left<a (t) \right>_{\text{coh}}$, assuming that $\left< a(t) \right> \approx \left< a(t) \right>_N$ is a reasonable approximation, we would conclude that the wavepacket would respond with the bare mass at all times, as occurs at $t=0$: although the intraband term \eqref{E:AccIntraAll} alone predicts a response with the usual effective mass, the interband term \eqref{E:AccInterAll} would ``mask" this response and would result in a response with the bare mass at all times. However, once the amplitudes in bands $n \ne N$ are included through \eqref{E:AccelCoh}, the effective mass behaviour is ``revealed'' as $\left< a(t) \right> $ oscillates around the usual effective mass according to \cite{Duque12} 
\begin{equation}
  \left< a(t) \right> \approx \int dk |c_N(\kappa)|^2 \bigg( \frac{F}{m_{N}^{\ast}(k)} +\\
    \frac{2F}{m_o^2} \sum_{n \ne N} \frac{E_{nN}(k)}{(E_{nN}(\kappa))^2}\text{Re} \left[ p_{Nn}(k)p_{nN}(\kappa) e^{i\gamma_{Nn}(\kappa,t)}\right] \bigg).
\end{equation}
Here $\text{Re}[\cdot]$ denotes the real part of the argument and $\gamma_{Nn}(\kappa,t) \equiv \gamma_{N}(\kappa,t) - \gamma_{n}(\kappa,t)$.

An interpretation of this result is that the usual effective mass behaviour comes from the \emph{intraband} part of the interaction (associated with $\hat{x}_i$) between Bloch states belonging to the \emph{same band} $N$. The initial bare mass response and the oscillatory behaviour around the usual effective mass come from the combined contributions of the \emph{interband} part of the interaction (associated with $\hat{x}_e$) between Bloch states of \emph{same band} $N$ and the effect of the \emph{coherence} created between the amplitude in band $N$ and its neighbours. This is of course consistent with the picture in terms of modified Bloch states, where these states are built from a coherent superposition of usual Bloch states.
